\documentclass[11pt,a4paper]{article}

\usepackage{amsfonts}
\usepackage{amsthm}

\def\N{\mathbb N}
\def\Z{\mathbb Z}
\def\C{\mathbb C}
\def\R{\mathbb R}
\def\dom{\mathrm {dom}\,}
\def\LL{\mathrm L}

\newcommand{\la}{\langle }
\newcommand{\ra}{\rangle }
\newcommand{\veps}{\varepsilon}
\newcommand{\im}{{\rm Im \hspace{0.05cm}}}

\newtheorem{teor2}{Theorem}

\newtheorem{lema2}{Lemma}

\theoremstyle{definition}
\newtheorem{exem2}{Example}

\theoremstyle{definition}

\begin{document}

 \title{Self-adjoint extensions of Coulomb systems in 1,2 and 3 dimensions}
\author{C\'esar R. de Oliveira {\small and} Alessandra A. Verri\\
\vspace{-0.6cm}
\small
\it Departamento de Matem\'{a}tica -- UFSCar, \small \it S\~{a}o Carlos, SP,
13560-970
Brazil\\ \\}
\date{\today}

\maketitle

\begin{abstract} We study the nonrelativistic quantum Coulomb hamiltonian (i.e., inverse of distance
potential) in
$\R^n$,
$n=1,2,3$. We characterize their self-adjoint extensions and, in the unidimensional case, present a discussion of
controversies in the literature,  particularly the question of the permeability of the origin. Potentials given by
fundamental solutions of Laplace equation are also briefly considered.
\end{abstract}

\section{Introduction} 

In principle the unidimensional ({\sc1D}) hydrogen atom is a simplification of the three-dimensional
({\sc3D}) model which has been invoked in theoretical and numerical studies \cite{JSS,DKS,LCO3}; note that
Cole and Cohen
\cite{CC} and Wong et al.\ \cite{Wong} have
reported some experimental evidence for the {\sc1D} hydrogen atom.  In a particular situation the
{\sc1D} eigenvalues coincide with the well-known
eigenvalues of the {\sc3D} hydrogen model, as discussed ahead.

Apparently the  {\sc1D} hydrogen atom was first considered in 1928  by Vrkljan \cite{VRs}. However, it was a work of
Loudon \cite{Lou} published in 1959,  whose potential model is
\[
V_C(x)= -\frac{\kappa}{|x|},\quad \kappa>0,
\] that increased attention to the subject which has become
interesting and quite controversial. We refer to $V_C$ as {\it the Coulomb potential}.

Loudon stated that the {\sc1D} hydrogen atom was twofold degenerate, having
even and odd eigenfunctions for each
eigenvalue, except for the (even) ground state having infinite binding
energy. Typically {\sc1D} systems have no
degenerate eigenvalues, and Loudon justified the double degeneracy as a
consequence of the singular atomic
potential. Andrews \cite{And1} questioned the existence of a ground
state with infinite binding energy. Ten
years later Haines and Roberts \cite{HR} revised Loudon's work and
obtained that their even wave functions,
with continuous eigenvalues, were complementary to odd functions, but
such results were criticized by
Andrews \cite{And2}, who did not accept the continuous eigenvalues.
Gomes and Zimerman \cite{GZ} argued
that the even states with finite energy should be excluded. Spector and
Lee \cite{SL} presented a
relativistic treatment that removed the problem of infinite binding
energy of the ground state. Several other
works \cite{DPST,BKB,NVS,LC,OL,FLM,XDD,LL,LCO6} (see also references therein)
have discussed this apparently
simple problem.

In this work we advocate that the roots of such controversies is a lack of sufficient mathematical care in some
papers: in {\sc 1D} the Coulomb singularity is so severe that it is not a trivial problem to assign boundary
conditions at the origin. The main question is how to properly define the self-adjoint realization(s) of
\[
\dot H=-\frac{\hbar^2}{2m}\Delta + V_C(x), \quad \dom \dot H = C_0^\infty(\R\setminus\{0\}).
\]The domain choice is  because functions $\psi$ in $C_0^\infty(\R\setminus\{0\})$ are kept far enough from the origin
(i.e., zero does not belong to their support), and so $\dot H\psi$ is well defined. $\dot H$ is hermitian but not
self-adjoint, and it turns out that it has deficiency index
$n_+=2=n_-$ and so an  infinite family of self-adjoint extensions (see Section~\ref{SAEsection1D}). Although such
extensions appear in
\cite{FLM}, details of how they were obtained are missing; in Section~\ref{SAEsection1D} we find such extensions by
another approach, that is, we use a (modified) boundary form as discussed in \cite{CRdO}. These extensions are the
candidates for the energy operator of the {\sc1D} hydrogen atom.

With these extensions at hand, we discuss the question of permeability of the origin, that is, whether in {\sc1D} the
Coulomb singularity acts as barrier that allows the electron to pass through it or not. This is one of the important
questions considered in the literature. It is found that the permeability depends on the self-adjoint extension and we
present explicit examples of both behaviors.

In Sections~\ref{SAEsection3D} and \ref{SAEsection2D} we present the self-adjoint extensions of $\dot H$ in 3 and 2
dimensions, respectively. It is well-known that in 3D the operator $H$, with the same action as $\dot H$ but domain
$C_0^\infty(\R^3)$, is essentially self-adjoint, that is, it has just one self-adjoint extension, whose domain is the
Sobolev space
$\mathcal{H}^2(\R^3)$ (this is known as Kato-Rellich Theorem; see ahead). However, if the origin is  removed and the
initial  domain
$C_0^\infty(\R^3\setminus\{0\})$ is considered, then also in {\sc3D} there are infinitely many self-adjoint extensions.
Note that in both {\sc1D} and {\sc2D} the origin must be removed in order to get well-defined initial operators $\dot
H$.

It  is sometimes assumed that the right potential describing the coulombian interaction is given by the fundamental
solutions of Laplace equation, that is,
\[
V_1(x) = \kappa |x|,\quad V_2(x)= \kappa \ln |x|,\quad V_3(x)=-\frac{\kappa}{|x|},
\]in {\sc 1D}, {\sc 2D}  and {\sc 3D}, respectively. For example, in the statistical mechanics of the
Coulomb gas in {\sc 2D} the potential $V_2$ is often considered,  instead of $V_3$, and the so-called Kosterlitz-Thouless
transition is  obtained. So  in Section~\ref{PotLaplacesection}  we consider the Schr\"odinger
operator with such potentials and argue  that they are always essentially self-adjoint (in suitable domains),
independently of dimension. Finally, the conclusions are reported in Section~\ref{ConclusionsSection}. 

A
notational detail: the dot in $\dot H$ means that the origin has ben removed from the domain of the initial hermitian
operator; e.g., in {\sc 1D} $\dom \dot H$ is $C_0^\infty(\R\setminus\{0\})$ and so on.

\section{Self-adjoint extensions: {\sc 3D}}\label{SAEsection3D}
The initial hermitian
operator modelling the nonrelativistic quantum {\sc 3D} hydrogen atom is (write $r=|x|,\theta,\varphi$ for the
spherical coordinates)
\[
H=-\frac{\hbar^2}{2m}\Delta + V_C(x), \quad \dom H = C_0^\infty(\R^3)\subset \LL^2(\R^3),
\]which is well defined since for $\psi\in\dom H$
\[
\|V_C\psi\|^2 = \kappa^2\int_{\R^3} \frac{|\psi(x)|^2}{|x|^2} dx =\kappa^2 \int_0^\infty dr \int_{0}^{\pi} d\theta
\sin\theta 
\int_0^{2\pi}d\varphi |\psi(r,\theta,\varphi)|^2 <\infty.
\]

The Kato-Rellich Theorem \cite{RS2,deOIST} applies to this case, since $V_C\in \LL^2(\R^3)+\LL^\infty(\R^3)$, and $H$
has just one self-adjoint extension  whose domain is the Sobolev space $\mathcal H^2(\R^3)$. Thus, the Schr\"{o}dinger
operator is well established in this case, so the quantum dynamics, and this is the standard operator discussed in 
textbooks on quantum mechanics (usually with less mathematical details). 

It is worth mentioning that in $\R^n$,
$n\ge4$, Kato-Rellich Theorem implies unique self-adjointness for potentials $V\in \LL^p(\R^n)+\LL^\infty(\R^n)$ with
$p>n/2$; so in dimensions $n\ge4$ the Schr\"odinger operators $H$ with potential $V_C$ and domain $C_0^\infty(\R^n)$
are always essentially self-adjoint.

However, in {\sc 1D} and {\sc 2D} the condition $\|V_C\psi\|^2<\infty$ for all $\psi\in C_0^\infty(\R^3)$
requires
$\psi(0)=0$ for $x$ in a neighbourhood of the origin, that is,
$\psi\in C_0^\infty(\R^n\setminus\{0\})$, $n=1,2$. The self-adjoint extensions in such cases will be discussed in other
sections; for question of comparison with other dimensions, now we consider what happens if the origin is removed also
in
$\R^3$, that is, if the initial hermitian operator is
\[
\dot H=-\frac{\hbar^2}{2m}\Delta + V_C(x), \quad \dom \dot H = C_0^\infty(\R^3\setminus\{0\}).
\]

Write $\xi=(\theta,\varphi)$ for the angular variables in the unit sphere $S^2$ and $d\xi=\sin\theta \,d\theta
d\varphi$. Let
$\mathcal D$ denote the set of linear combinations of products $f(r)w(\xi)$ with $f\in C_0^\infty(0,\infty)\subset
\LL^2((0,\infty),r^2dr)$ and
$w\in \LL^2(S^2,d\xi)$. Due to the decomposition in spherical coordinates
\[
\LL^2(\R^3) = \LL^2((0,\infty),r^2dr) \otimes \LL^2(S^2,d\xi),
\] $\mathcal D$ is a dense set in $\LL^2(\R^3)$. For functions $\phi(r,\xi)=f(r)w(\xi)\in\mathcal D$ the
operator
$\dot H$ takes the form 
\[
\dot H f(r)w(\xi) = \left[-\frac{\hbar^2}{2m} \left(\partial_r^2 + \frac2r \partial_r  \right)f(r)-\frac\kappa r f(r) 
\right]w(\xi) + \frac{\hbar^2}{2m}\frac{f(r)}{r^2}\mathcal Bw(\xi),
\]where $\mathcal B$ is the Laplace-Beltrami operator \cite{Msh}
\[
(\mathcal Bw)(\xi) = -\frac1{\sin\theta}\left[ \partial_\theta \left(\sin\theta \partial_\theta w  \right)
+\frac1{\sin\theta} \partial^2_\varphi w\right]
\]acting in $\LL(S^2,d\xi).$ $\mathcal B$ with domain $C_0^\infty(S^2)$ is essentially self-adjoint and its
eigenfunctions are the spherical harmonics $Y_{l,m}(\xi)$, which constitute an orthonormal basis of $\LL^2(S^2,d\xi)$;
recall that
\[
(\mathcal BY_{l,m})(\xi) = l(l+1) Y_{l,m}(\xi),\quad l\in\N,\, -l\le m\le l.
\]

Denote by $\mathcal J_l$ the subspace spanned by $\{Y_{l,m}:  -l\le m\le l \}$, that is, the subspace corresponding
to the eigenvalue $l(l+1)$ and $\LL_l :=\LL^2((0,\infty),r^2dr) \otimes \mathcal J_l$; thus
\[
\LL^2(\R^3) = \LL^2((0,\infty),r^2dr) \otimes \LL^2(S^2,d\xi) = \bigoplus_{l=0}^\infty \LL_l.
\]
If $I_l$ is the identity operator on $\mathcal J_l$, the restriction of $\dot H$ to $\mathcal D_l = \mathcal D\cap
\LL_l$ is given by $\left. \dot H\right|_{\mathcal D_l} = \dot H_l \otimes I_l,$ with
\[
\dot H_l = -\frac{\hbar^2}{2m} \left(\frac{d^2}{dr^2} + \frac2r \frac{d}{dr} 
-\frac{l(l+1)}{r^2}\right)-\frac\kappa r,  
\]and our task is reduced to finding the self-adjoint extensions of $\dot H_l$ with domain $C_0^\infty(0,\infty)$. It
is convenient to introduce the unitary transformation $U:\LL^2((0,\infty),r^2\,dr)\to \LL^2(0,\infty)$,
$(U\phi)(r)=r\phi(r)$, which maps $C_0^\infty(0,\infty)$ to itself and
\[
h_l:= U\dot H_l U^{-1} = -\frac{\hbar^2}{2m} \left(\frac{d^2}{dr^2} -\frac{l(l+1)}{r^2}\right)-\frac\kappa r,
\] with domain $\dom H_l = C_0^\infty(0,\infty)$. Standard arguments gives that the adjoint $h_l^*$ has the same
action as $h_l$ but with domain 
\[
\dom h_l^* = \left\{\phi\in \LL^2(0,\infty): \phi,\phi'\in \mathrm{AC}(0,\infty), h_l^*\phi\in \LL^2(0,\infty) 
\right\}. 
\] If $\Omega$ is an open subset of $\R$, $\mathrm{AC}(\Omega)$ indicates the set of absolutely continuous functions
in every bounded and closed subinterval of $\Omega$. 

By adapting the analysis  of the free hamiltonian in
$\R^3$ with the origin removed, which was performed in
\cite{AJS}, one proves the following result:

\begin{teor2}\label{teorR3partes}
$h_0$ has deficiency indices equal to 1, while $h_l$, $l\ne0$, is essentially self-adjoint.
\end{teor2}

Hence, for $l\ne0$ the unique self-adjoint extension of $h_l$ is $h_l^*$, while $h_0$ has infinitely many self-adjoint
extensions. In order to find such extensions in case $l=0$, we will make use of the following lemma
\cite{Kuras,Kuras2,Moshinsky}, whose proof we adapt and reproduce.

\begin{lema2}\label{lemah0} If $\phi\in\dom h_0^*$, then the lateral limits $\phi(0^+):=
\lim_{r\to 0^+}\phi(r)$ and
\[\tilde\phi(0^+)  := \lim_{r\to0^+} \left( \phi'(r) + \frac{2m\kappa}{\hbar^2} \phi(r) \ln (\kappa r) \right)
\]  exist  (and are finite).
\end{lema2}
\proof For $\phi\in\dom h_0^*$ one has 
\[  - h_0^*\phi = \frac{\hbar^2}{2m} \frac{d^2\phi}{dr^2} +\frac\kappa r\phi:=u \in\LL^2(0,\infty),
\]and one can write $\phi = \phi_1+\phi_2$ with  $\frac{\hbar^2}{2m}\phi_1''=u$, $\phi_1(0^+)=0$ and
$\frac{\hbar^2}{2m}\phi_2''+\kappa/r \phi=0.$ Since $\phi_j\in {\mathcal H}^2(\varepsilon,\infty)$,
$j=1,2$, for all $\veps>0$, and $u \in\LL^2$, it follows that these functions are of class $C^1(0,\infty)$. 

Consider an interval $[r,c]$,  $0<r<c<\infty$. Since
\[
\phi_1'(r)-\phi_1'(c) = \frac{2m}{\hbar^2}\int_r^c u(s)\,ds,
\]
$\phi_1'(r)$ has a lateral limit
\[
\phi_1'(0^+)= \phi'(c)+ \frac{2m}{\hbar^2}\int_0^c u(s)\,ds.
\] On integrating successively twice over the interval $[r,c]$ one gets
\[
\phi_2'(c) -\phi_2'(r) = -\frac{2m\kappa}{\hbar^2}\int_r^c \frac{\phi(s)}{s}\,ds,
\]and then
\[
\phi_2(r) = \phi_2(c) - (c-r)\phi_2'(c) -\frac{2m\kappa}{\hbar^2}  \int_r^c dv \int_v^c
ds\frac{\phi(s)}{s}
\]
\[
= \phi_2(c) - (c-r)\phi_2'(c) -\frac{2m\kappa}{\hbar^2}  \int_r^c ds\phi(s)\,\frac{s-r}{s},
\]and since $0\le(s-r)/s<1$, by dominate convergence the last integral converges to $\int_0^1 \phi(s)$ as $r\to
0^+$. Therefore
$\phi_2(0^+)$ exists and
\[
\phi_2(0^+)  =\phi_2(c) - c\phi_2'(c) - \frac{2m\kappa}{\hbar^2} \int_0^c\phi(s)\,ds.
\]
Now,\[
\left| \phi_2(r)-\phi_2(0^+) \right| \le r|\phi_2'(c)| + \frac{2m\kappa}{\hbar^2} \int_0^r |\phi(s)|\,ds +
\frac{2m\kappa}{\hbar^2} r \int_r^c ds\, \frac{|\phi(s)|}{s}.
\]Taking into account that $\phi$ is bounded, say $|\phi(r)|\le C$, $\forall r$, Cauchy-Schwarz in $\LL^2$
implies
\[
\int_0^r\,|\phi(s)|ds = \int_0^r\,1\,|\phi(s)|ds \le C \sqrt{r},
\] and so, for $r$ small enough and fixing $c=1$, 
\[
\int_r^c ds \frac{\phi(s)}{s} \le C \left(c|\ln c|+r|\ln  r|\right)\le \tilde C \sqrt r,
\] for some constant $\tilde C$. Such inequalities imply $\phi(r) = \phi(0^+) +
O(\sqrt r)$, and on substituting this into 
\[
\phi'(r) =\phi'(1) + \frac{2m\kappa}{\hbar^2}\int_r^1 \frac{\phi(s)}{s}\,ds
\] (recall that $\phi_1'(0^+)$ is finite) it is found that there is $b$ so that, as $r\to 0^+$, 
\[
\phi'(r) = \phi'(1) -
\frac{2m\kappa}{\hbar^2}\phi(0^+) \ln(\kappa r) + b + o(1);
\]thus, the derivative $\phi'$ has a a logarithmic divergence as $r\to0$ and the statement in the lemma also
follows.
\endproof

 For $\phi,\psi\in \dom h_0^*$ integration by parts gives
\[
\la h_0^*\psi,\phi\ra - \la \psi,h_0^*\phi\ra =\Gamma(\psi,\phi),
\]where 
\[
\Gamma(\psi,\phi):=-\frac{\hbar^2}{2m}\;\lim_{r\to 0^+}\left(\psi(r)\overline{\phi'(r)}
-\psi'(r)\overline{\phi(r)}
\right)
\] is called a {\it boundary form} for $h_0$ \cite{CRdO}. Note that although $\Gamma(\psi,\phi)$ is finite, the lateral
limit $\phi'(0^+)$ can diverge; however, by Lemma~\ref{lemah0} it is readily verified that
\[
\Gamma(\psi,\phi)=-\frac{\hbar^2}{2m}\left(\psi(0^+)\overline{\tilde\phi(0^+)}
-\tilde\psi(0^+)\overline{\phi(0)}
\right)
\]and now all lateral limits are finite. The self-adjoint extensions of $h_0$ are restrictions of $h_0^*$ to suitable
subspaces $D$ so that $\dom h_0\subset D\subset \dom h_0^*$ and $\Gamma|_D=0$, that is, $\Gamma(\psi,\phi)=0$ for all
$\phi,\psi\in D$. 

We introduce the unidimensional vector spaces $X=\{\rho_1(\psi):=\psi(0^+)+i\tilde\psi(0^+): \psi\in\dom h_0^*  \}$ and
$Y=\{\rho_2(\psi):=\psi(0^+)-i\tilde\psi(0^+): \psi\in\dom h_0^*  \}$ and note that
\[
\frac{4mi}{\hbar^2} \Gamma(\psi,\phi) = \la \rho_1(\psi),\rho_1(\phi)\ra_X - \la \rho_2(\psi),\rho_2(\phi)\ra_Y ,
\]with inner products in $X,Y$, as indicated. Hence, the subspaces $D$ for which $\Gamma$ vanishes are related to maps
that preserve inner products, that is, unitary maps from $X$ to $Y$ (see details in \cite{CRdO}), and since these
vector spaces are unidimensional such maps are multiplication by the complex numbers
$e^{i\theta}$, $0\le\theta<2\pi$. Therefore, for each $\theta$ a self-adjoint extension of $h_0$ is characterized by
the functions $\psi\in\dom h_0^*$ so that $\rho_2(\psi)=e^{i\theta}\rho_1(\psi)$, that is,
\[
(1-e^{i\theta})\psi(0^+) = i (1+e^{i\theta})\tilde\psi(0^+).
\]If $\theta\ne0$ this condition reduces to
\[
\psi(0^+) = \lambda\tilde\psi(0^+),\quad \lambda = i\frac{1+e^{i\theta}}{1-e^{i\theta}}\in \R,
\]and writing $\lambda=\infty$ in case $\theta=0$, the desired self-adjoint extensions $h_0^\lambda$ are 
described by
\[
\dom h_0^\lambda = \left\{\psi\in\dom h_0^*:  \psi(0^+) = \lambda\tilde\psi(0^+) \right\},\quad \lambda\in \R\cup
\{\infty\},
\]and $h_0^\lambda\psi = h_0^*\psi.$ The Dirichlet boundary condition corresponds to $\lambda=0$. With such results at
hand, we have

\begin{teor2}\label{teorExtAA3D}
The self-adjoint extensions of $\dot H$ in {\sc 3D} are
\[
H^\lambda = \left( U^{-1} h_0^\lambda U \otimes I_0 \right) \bigoplus_{l=1}^\infty \left( U^{-1} h_l^* U \otimes
I_l \right), \quad \lambda\in \R\cup \{\infty\}.
\]
\end{teor2}

This should be compared with the case without removing the origin, for which there is just one self-adjoint extension.
The eigenvalue equation for the Dirichlet case $\lambda=0$ can be exactly solved in terms of Whittaker functions, and
the negative eigenvalues are
\[
E_n^0 = -\frac{\kappa^2 m}{2\hbar^2}\frac{1}{n^2},\quad n=1,2,3,\cdots,
\]each one with multiplicity $n^2$. For $\lambda\ne0$ the manipulations become more involved and numerical procedures
must be employed to find roots of implicit functions, and so the eigenvalues.

\section{Self-adjoint extensions: {\sc 2D}}\label{SAEsection2D}
Although our main interest is in the {\sc 1D} and {\sc 3D} cases, we say something about the Coulomb system in {\sc
2D}. As already mentioned,  the origin must be excluded from the domain of the Coulomb potential in {\sc 2D} and
the initial hermitian operator is
\[
\dot H=-\frac{\hbar^2}{2m}\Delta + V_C(x), \quad \dom \dot H = C_0^\infty(\R^2\setminus\{0\}).
\] To find its self-adjoint extensions, introduce polar coordinates $(r,\varphi)$ so that
\[
\LL^2(\R^2) = \LL^2((0,\infty),rdr) \otimes \LL^2(S^1,d\varphi),
\] ($S^1$ is the usual unit circle in $\R^2$) and the set ${\mathcal D}$ of linear combinations of the products
$f(r)g(\varphi)$,
$f\in C_0^\infty(0,\infty)$ and $g\in C_0^\infty(S^1)$, is dense in $\LL^2(\R^2)$. Now
\[
\dot H f(r)g(\varphi) = \left[-\frac{\hbar^2}{2m} \left(\partial_r^2 + \frac1r \partial_r  \right)f(r)-\frac\kappa r
f(r) 
\right]g(\varphi) - \frac{\hbar^2}{2m}\frac{f(r)}{r^2}\mathcal Bg(\varphi),
\]
where $\mathcal B=\partial_\varphi^2$ is the Laplace-Beltrami operator acting in $\LL^2(S^1,d\varphi)$. This operator
with domain $C_0^\infty(S^1)$ is essentially self-adjoint, its eigenvectors
$g_l(\varphi)=e^{il\varphi}/\sqrt{2\pi}$ constitute an orthonormal basis of $\LL^2(S^1,d\varphi)$ and
\[
(\mathcal Bg_l)(\varphi) = -l^2 g_l(\varphi),\quad l\in\Z.
\]

Let $[g_l]$ denote the subspace spanned by $g_l$ and $\LL_l=\LL^2((0,\infty),rdr)\otimes[g_l]$; thus
\[
\LL^2(\R^2) = \bigoplus_{l\in\Z} \LL_l,
\]and if $I_l$ is the identity operator on $[g_l]$, the restriction
of $\dot H$ to $\mathcal D_l = \mathcal D\cap
\LL_l$ is given by $\left. \dot H\right|_{\mathcal D_l} = \dot H_l \otimes I_l,$ with
\[
 \dot  H_l = -\frac{\hbar^2}{2m} \left(\partial_r^2 + \frac1r \partial_r -\frac{l^2}{r^2}
\right)-\frac\kappa r,
\] with domain $C_0^\infty(0,\infty)$, and the question is to find the self-adjoint extensions of such restrictions.
By using the unitary operator $U: \LL^2((0,\infty),rdr)\to  \LL^2(0,\infty) $, $(U\phi)(r) = r^{1/2}\phi(r)$, one has
\[
h_l:= U\dot H_lU^{-1}= -\frac{\hbar^2}{2m} \left(\partial_r^2 +  \left( \frac14-l^2
\right)\frac{1}{r^2}
\right)-\frac\kappa r
\]with $\dom h_l= C_0^\infty(0,\infty)$ (since this set is invariant under $U$). By standard results it follows
that the adjoint
$h_l^*$ has the same action as $h_l$ but with domain
\[
\dom h_l^* = \left\{\phi\in \LL^2(0,\infty): \phi,\phi'\in \mathrm{AC}(0,\infty), h_l^*\phi\in\LL^2(0,\infty)   
\right\}.
\]

\begin{teor2}\label{teor2D}
The operators $h_l$ are essentially self-adjoint if, and only if, $l\ne0$, whereas $h_0$ has deficiency indices equal
to one.
\end{teor2}
\proof Weyl's limit point-limit circle criterion will be used \cite{RS2}. Thus we consider the solutions of $h_l^* \phi
= i
\phi$, that is,
\[
-\frac{\hbar^2}{2m} \phi'' - 
\left[ \frac{\hbar^2}{2m} \left( \frac{1}{4} - l^2 \right)  \frac{1}{r^2} 
+ \frac{\kappa}{r} + i\right] \phi = 0,\quad \phi\in \dom h_l^*.
\]
By writing $p =   \frac{2m\kappa}{\hbar^2}$,
$q =   \frac{2mi}{\hbar^2}$ and performing the change of variable 
$y = (-4 q)^{1/2} r$, one gets
\[
\phi''+ \left[ \left( \frac{1}{4} - l^2 \right) \frac{1}{y^2} + 
\frac{\tau}{y}-\frac{1}{4} \right] \phi = 0,
\]
with $\tau = p /(- 4 q)^{1/2}$.
This equation has two linearly independent solutions given by the Whittaker functions
\cite{Abro,Grad}
\[\phi_1 (y) = {\cal M}_{\tau, |l|}(y) \quad  \hbox{e} \quad 
\phi_2 (y) = {\cal W}_{\tau, |l|}(y),
\]whose asymptotic behaviors as $|y| \to \infty$ are
\[\phi_1 (y)  \sim e^{y/2} (-y)^{-\tau}
\quad  \hbox{e} \quad 
\phi_2(y) \sim  e^{-y/2} y^{\tau}.
\]

Since there is no $c \in \mathbb R$ so that $\phi_2 \in \LL^2(c, \infty)$, 
it follows that $h_l$ is limit point at $\infty$. Note that the above asymptotic behaviors as $|y| \to \infty$
do not depend on  $l$; however, at the origin we need to separate the cases $l=0$ and $l \neq 0$.

For $l= 0$ \cite{Abro,Grad},
\[
\phi_1(0^+) = 0 \quad  \hbox{and} \quad  \phi_2(0^+)=0.
\]
In this case there is $c>0$ so that $\phi_1,\phi_2 \in \LL^2(0, c)$ and $h_0$ is limit circle at 0. Therefore,
$h_0$ is not essentially self-adjoint but has deficiency indices equal to 1.

For $l \neq 0$,
$\phi_1 (0^+) = 0$ while
$\phi_2 (y)$ diverges as  
\[
 \sum_{k=0}^{2 |l|-1} \frac{\Gamma(2 |l| - k)}{k!}\Gamma(k - |l|- \tau + 1/2) (-y)^{-2|l|+k},
\]
for $y \to 0^+$ (here $\Gamma$ denotes the well-known Gamma function).
Therefore, $h_l$ ($l \neq 0$)  is limit point at $0$. By Weyl criterion, $h_l$ is essentially self-adjoint if $l
\neq 0$.
\endproof

Since $h_l$, $l\ne0$, is essentially self-adjoint, its unique self-adjoint extension is exactly $h_l^*$.  According to
the proof of Theorem~\ref{teor2D}, the deficiency subspace
$K_-(h_0)$ is spanned by
$\phi_-(r)={\cal M}_{\tau, 0}((-4 q)^{1/2}r)$ and the deficiency subspace $K_+(h_0)$ spanned by
$\phi_+(r)=\overline{\phi_-(r)}$. The von Neumann theory of self-adjoint extensions \cite{RS1,deOIST} characterizes
them by unitary maps from, say, $K_-$ to $K_+$, and since such subspaces are unidimensional these unitary maps are just
multiplication by $e^{i\theta}$, $0\le
\theta<2\pi$. Thus there is a self-adjoint extension $h_0^\theta$ of $h_0$, for each $\theta$, and if $\overline{h_0}$
denotes the closure of $h_0$, it is given by
\[
\dom h_0^\theta = \left\{ \psi+c \left(\phi_+ - e^{i\theta}\phi_-  \right): \psi\in\dom\overline{h_0}, \, c\in\C 
\right\},
\]
\[
h_0^\theta \psi = h_0^*\psi,\quad \psi\in\dom h_0^\theta.
\] In summary:

\begin{teor2}\label{teorExtAA2D}
The self-adjoint extensions of $\dot H$ in {\sc 2D} are
\[
H^\theta = \left( U^{-1} h_0^\theta U \otimes I_0 \right) \bigoplus_{l\in\Z,l\ne0}^\infty \left( U^{-1} h_l^* U \otimes
I_l \right), \quad \theta\in [0,2\pi).
\]
\end{teor2}

\section{Hydrogen atom: {\sc 1D}}\label{SAEsection1D}
In the unidimensional case the initial hermitian operator is
\[
\dot H = -\frac{\hbar^2}{2m} \frac {d^2}{dx^2} - \frac{\kappa}{|x|},\quad \dom \dot H = C_0^\infty(\R\setminus\{0\}),
\]and the origin naturally decomposes the space 
\[
\LL^2(\R\setminus\{0\}) = \LL^2(-\infty,0) \oplus \LL^2(0,\infty),
\]
and also the domain of $\dot H$ into $C_0^\infty(-\infty,0)$ and
$C_0^\infty(0,\infty)$; let $\dot H_+$ and $\dot H_-$ denote the restriction of $\dot H$ to these subspaces,
respectively. Thus, we have 
\[
\dot H = \dot H_- \oplus \dot H_+.
\]From the physical point of view, an important question is about the behavior of the system at the origin, e.g., is
it impermeable, so that the system actually decomposes into a right one and a left one, or is it permeable? In the
latter possibility, what do happen with wavefunctions at the transition point (the origin)? As discussed ahead, there
are  plenty of possibilities, due to infinitely many self-adjoint extensions.

\subsection{Self-adjoint extensions}\label{subsectSae}
The adjoint operator $\dot H_+^*$ has the same action as $\dot H$ but with domain
\[
\dom \dot H_+^* = \left\{\phi\in \LL^2(0,\infty): \phi,\phi'\in \mathrm{AC}(0,\infty), \,H_+^*\phi\in  \LL^2(0,\infty) 
\right\},
\]and an analogous expression for $\dot H_-^*$ and its domain. 

To find the deficiency subspace
$K _+(\dot H_+)$ we look for solutions of
\[
- \frac{\hbar^2}{2m}  \phi '' + \left(- \frac{\kappa}{x} + i \right) \phi = 0
\]
that belong to $\dom \dot H_+^*$.
Write  $p =   \frac{2m \kappa}{\hbar^2}$, $q = -   \frac{2 m i}{\hbar^2}$ and perform the change of variable
$y = (-4 q)^{1/2} x$, so that this equation takes the form
\[ \phi''(y) + \left( \frac{\tau}{y} - \frac{1}{4} \right) \phi(y) = 0,
\]
with $\tau = p/(-4 q)^{1/2}$, which has exactly two linearly independent solutions \cite{Abro,Grad}
\[\phi_{1+} (y) = {\cal W}_{\tau, 1/2} (y) ,\quad
\phi_{2+} (y) = {\cal M}_{\tau, 1/2} (y).
\]
These solutions have finite limites as $y \to 0$. The asymptotic behaviors for $|y| \rightarrow \infty$ are
\[\phi_{1+} (y) \sim y^{\tau} e^{-y/2} \quad \mathrm{and}\quad  
\phi_{2+} (y) \sim (-y)^{-\tau} e^{y/2}.
\]

Hence, in the original variable the deficiency subspace $K_+(\dot H_+)$
is unidimensional and spanned by $\phi_{1+}(x)= {\cal W}_{\tau, 1/2}((-4q)^{1/2}x)$. A similar analysis implies that
$K_- (\dot H_+)$ is also unidimensional and spanned by $ \overline{\phi_{1+}(x)}$. Therefore $\dot H_+$ has
both deficiency indices equal to 1.

Similarly one finds that the deficiency subspaces $K_+(\dot H_-)$ and $K_-(\dot H_-)$ of  $\dot H_-$ are spanned,
respectively, by
$\phi_{1-}(x) = {\cal W}_{\tau, 1/2} ((-4q)^{1/2}|x|)$ and 
$ \overline{\phi_{1-}(x)}$.
Therefore $\dot H_-$ also has both deficiency indices equal to 1.

Now, due to the above results, the adjoint operator $\dot H^*$ has the same action as $\dot H$ but domain (write
$\mathcal H=\LL^2 ( \R\setminus \{0\})$)
\[\dom H^*  =  
\left\{ \phi \in \mathcal H  : \phi, \phi' \in \mathrm{AC} (\R\setminus \{0\} ),
- \frac{\hbar^2}{2m}  \phi'' - \frac{\kappa}{|x|} \phi  \in \mathcal H \right\},
\]
and both deficiency subspaces $K_\pm (\dot H)$ have dimension $2$; $K_+ (\dot H)$  is spanned by
\[
\psi_1 (x)  = \left \{
\begin{array}{rll}
\phi_{1+}(x)   &  \mathrm{if}  &  x > 0 \\
        0                  &  \mathrm{if}  &  x < 0
\end{array} \right. \quad \mathrm{and}\quad  
\psi_2 (x)  =  \left \{
\begin{array}{rll}
        0                   &  \mathrm{if}  &  x > 0 \\
\phi_{1-}(x)  &  \mathrm{if}  &  x < 0
\end{array} \right.,
\]and $K_- (\dot H)$ by $\overline{\psi_1}$ and $\overline{\psi_2}$.

Therefore,
$\dot H$ has both deficiency indices equal to $2$ and it has infinitely many self-adjoint extensions.
A boundary form will be used in order to get such extensions. The following lemma will be needed, whose proof is
similar to the proof of Lemma~\ref{lemah0} (note
the difference of signs in the definitions of $ \tilde{\phi}(0^+)$ and $ \tilde{\phi}(0^-)$ below).

\begin{lema2}\label{lila}
If $\phi \in \dom  \dot H^*$, then the lateral limits $\phi(0^\pm):=
\lim_{x\to 0^\pm}\phi(x)$ and
\[
  \tilde{\phi} (0^{\pm}) :=   \lim_{x \rightarrow 0^{\pm}} 
\left( \phi'(x) \pm \frac{2m \kappa}{\hbar^2} \phi(x) \ln( \pm \kappa x) \right)
\]
exist and are finite.
\end{lema2}

\

For $\psi,\phi\in\dom \dot H^*$ one has, upon integrating by parts,
\[
\la \dot H^*\psi,\phi\ra - \la \psi,\dot H^*\phi\ra =\Gamma(\psi,\phi),
\]where 
\[
-\frac{2m}{\hbar^2}\;\Gamma(\psi,\phi)=  
\]
\[=\left[\lim_{x\to 0^+}\left(
\psi(x) \overline{\phi'(x)} - \psi'(x) \overline{\phi(x)}\right) + \lim_{x\to 0^-}\left(
- \psi(x) \overline{\phi'(x)} + \psi'(x) \overline{\phi(x)}\right) 
\right]
\]
and, using  Lemma~\ref{lila}, straightforward computation gives 
\[\Gamma (\psi, \phi)   =    
\frac{-\hbar^2}{2m}\left( 
\psi(0^+) \overline{\tilde{\phi}(0^+)} - \tilde{\psi}(0^+) \overline{\phi(0^+)} -
\psi(0^-) \overline{\tilde{\phi}(0^-)} + \tilde{\psi}(0^-) \overline{\phi(0^-)}
\right),\]
and now each lateral limit is finite.

Introduce two linear maps
$\rho_1, \rho_2: \dom \dot H^* \to \C^2$:
\[ \rho_1(\psi) = \left(\begin{array}{ll}
\tilde{\psi}(0^+) + i \psi (0^+)\\
\tilde{\psi}(0^-) - i \psi (0^-)
          \end{array}\right)
\quad\mathrm{and}\quad
\rho_2(\psi) = \left(\begin{array}{ll}
\tilde{\psi}(0^+) - i \psi (0^+)\\
\tilde{\psi}(0^-) + i \psi (0^-)
          \end{array}\right),
\]
so that
\[
\langle \rho_1(\psi), \rho_1(\phi) \rangle_{\C^2} - 
\langle \rho_2(\psi), \rho_2(\phi) \rangle_{\C^2} =
-\frac{4m}{\hbar^2} i \,\Gamma (\psi, \phi),
\quad \forall \psi, \phi \in \dom \dot H^*.
\] 
As in Section~\ref{SAEsection3D}, the self-adjoint extensions of $\dot H$ are restrictions of $\dot H^*$ to
suitable subspaces $D$ so that $\dom \dot H\subset D\subset \dom \dot H^*$ and $\Gamma|_D=0$, that is,
$\Gamma(\psi,\phi)=0$ for all
$\phi,\psi\in D$ \cite{CRdO}.

Vanishing of $\Gamma$ on domains $D$ is equivalent to the preservation of the inner products in $\C^2$,
and so it corresponds to unitary  $2 \times 2$ matrices $\hat U$, and each of such matrices characterizes a
self-adjoint extension $\dot H_{\hat U}$ of $\dot H$, so that $\dom \dot H_{\hat U}$ is composed of  
$\psi \in \dom H^*$
so that $\rho_2(\psi) = \hat U \rho_1(\psi)$; also $\dot H_{\hat U}\psi = \dot H^*\psi$ for $\psi\in\dom H_{\hat U}$.

The condition $\rho_2(\psi) = \hat U \rho_1(\psi)$ is then written
\[
(I - \hat U) \left(\begin{array}{cc}
\tilde{\psi}(0^+) \\
\tilde{\psi}(0^-)
          \end{array}\right) =
- i (I + \hat U) \left(\begin{array}{cc}
-\psi(0^+) \\
\psi(0^-)
          \end{array}\right),
\]and we have explicitly got the boundary conditions characterizing the desired self-adjoint extensions.

In case $(I-\hat U)$ is invertible (similarly if $(I+\hat U)$ is invertible) it is possible to write the above boundary
conditions in the form
\[ 
\left(\begin{array}{cc}
\tilde{\psi}(0^+) \\
\tilde{\psi}(0^-)
          \end{array}\right)
= A
\left(\begin{array}{cc}
-\psi(0^+) \\
\psi(0^-)
          \end{array}\right),
\]
with $A = -i (I-\hat U)^{-1} (I+\hat U)$  being a $2 \times 2$ self-adjoint matrix.

In \cite{FLM} the authors have got this form for the self-adjoint extensions, and it was claimed that by allowing the
entries of $A$ taking infinity all self-adjoint extensions are found; we think it is a hard task to cover all
possibilities above (i.e., via $\hat U$) with this representation via self-adjoint matrices $A$.

Particular choices of the matrix $\hat U$ ($I$ is the identity matrix)
\[
\mathrm{a)}\;I,\qquad \mathrm{b)}\;-I,\qquad\mathrm{c)}\;\left(\begin{array}{cc}
0   &   1 \\
1   &     0
\end{array}\right),\qquad \mathrm{d)} \;\left(\begin{array}{cc}
0   &   -1 \\
-1   &     0
\end{array}\right),
\]impose, respectively, the boundary conditions: a) $\psi(0^-)=0=\psi(0^+)$ (Dirichlet); b)
$\tilde\psi(0^-)=0=\tilde\psi(0^+)$ (``Neumann''); c) $\psi(0^-)=\psi(0^+)$ and
$\tilde\psi(0^-)=\tilde\psi(0^+)$ (periodic); d)
$\psi(0^-)=-\psi(0^+)$ and $\tilde\psi(0^-)=-\tilde\psi(0^+)$ (antiperiodic).

Recall that the general form of a $2\times 2$ unitary matrix is
\[
\hat U = e^{i \theta}\left(
\begin{array}{cc}
a   &   - \overline{b} \\
b   &     \overline{a}
\end{array}\right),
\quad \theta \in [0, 2 \pi), \quad  a,b \in \mathbb C, 
|a|^2+ |b|^2 = 1.
\]This form will be used ahead.

\subsection{Negative eigenvalues}\label{subsectNE1} In this subsection we discuss the negative eigenvalues of some
self-adjoint extensions $\dot H_{\hat U}$. The main goal is to remark that the eigenvalues,  their multiplicities and 
parity of eigenfunctions depend on the boundary conditions. This becomes important since in the past
some authors have assumed particular hypotheses on the eigenfunctions (see references in the Introduction), but
without specifying the self-adjoint extension they were working with; this was the main source of controversies in the
studies of the unidimensional hydrogen atom.

Our first analysis is for $\hat U=I$, i.e.,   Dirichlet boundary condition,
and denote by $H_D$ the operator 
\[
H_D =   - \frac{\hbar^2}{2m}  \frac{d^2}{dx^2} - \frac{\kappa}{|x|}, \quad 
\dom H_D = \{\phi \in \dom \dot H^*:
\phi(0^+) = 0 = \phi(0^-) \}.
\]
As in \cite{FLM} we consider the Green function of $(H_D - E)^{-1}$, denoted by
$G(x,y)$, that is
\[
(H_D - E)^{-1} u(x)   =   \Theta(x) \int_0^x
G(x,y)  u(y) dy +  \Theta(-x) \int_x^0
G(x,y)  u(y) dy.
\] Here $\Theta(x)=1$ if $x>0$ and vanishes if $x<0$.

If $u \in\mathrm{rng} (H_D - E)$,
we search for solutions $\phi$ of 
\begin{equation}\label{equackernel1d}
(H_D - E) \phi = u,
\end{equation}
 and the discussion is for $(-\infty, 0)$ and $(0, \infty)$ separately. 

For $x \in (0, \infty)$, by the method of variation of parameters, one finds the solution 
\[
\phi(x) =  \phi_1(x) \int_{0}^{x}  - \frac{\phi_2(y) u(y)}{W_x(\phi_1, \phi_2)} dy  +
\phi_2(x) \int_{0}^{x}  \frac{\phi_1(y) u(y)}{W_x(\phi_1, \phi_2)}  dy,
\]
where $\phi_1$ and $\phi_2$ are independent solutions of the homogeneous equation 
$(H_D - E) \phi = 0$, that is,
\begin{equation}\label{autovalorE}
- \frac{h^2}{2m}  \phi '' - \left(\frac{\kappa}{x} + E \right) \phi = 0,
\end{equation}
and $W_x(\phi_1, \phi_2)$ the corresponding Wronskian.

Writing $p =  \frac{2m \kappa}{h^2}$, $q =  \frac{2 m E}{h^2}$, $\tau =  p/ (-4 q)^{1/2}$ and $z = (-4 q)^{1/2} x$,
the last equation takes the form
\[
 \phi '' + \left(\frac{\tau}{z} - \frac{1}{4} \right) \phi = 0,
\]
whose independent solutions are
\[
\phi_1 (z) = {\cal W}_{\tau, 1/2} (z) \quad  \mathrm{and} \quad 
\phi_2 (z) = {\cal M}_{\tau, 1/2} (z).
\]

Recall that ${\cal W}_{\tau, 1/2} (z) \sim e^{-z/2} z^\tau$ and 
${\cal M}_{\tau, 1/2} (z) \sim e^{z/2} (-z)^{-\tau}$, as $z \to \infty$.
In the original variable\[
W_x(\phi_1, \phi_2) = 
- \frac{(-4 q)^{1/2}}{\Gamma(1-\tau)},
\]
 and the unique solution satisfying $\phi(0^+)=0$ is
\begin{eqnarray*}
\phi(x) =
 \int_{0}^{x} \frac{\Gamma (1-\tau)}{(-4q)^{1/2}} 
&  \left({\cal W}_{\tau, 1/2}( (-4 q)^{1/2} x) \right. & 
{\cal M}_{\tau, 1/2}( (-4 q)^{1/2} y) - \\
&  {\cal M}_{\tau, 1/2}( (-4 q)^{1/2} x) &  \left.{\cal W}_{\tau, 1/2}( (-4 q)^{1/2} y)
\right) u(y) dy.
\end{eqnarray*}
 
Similarly, for $x \in (-\infty, 0)$, the unique solution satisfying $\phi(0^-)=0$ is
\begin{eqnarray*}
\phi(x) = \int_{x}^{0} \frac{\Gamma (1-\tau)}{(-4q)^{1/2}}
&\left({\cal W}_{\tau, 1/2}( (-4 q)^{1/2} |x|) \right.  & 
{\cal M}_{\tau, 1/2}( (-4 q)^{1/2} |y|) - \\
&{\cal M}_{\tau, 1/2}( (-4 q)^{1/2} |x|) & \left. {\cal W}_{\tau, 1/2}((-4 q)^{1/2}|y|)
\right) u(y) dy.
\end{eqnarray*}

Summing up, the Green function of the resolvent operator  ($H_D - E$)$^{-1}$  is given by 
\[
G(x,y)  = 
\Theta (xy) \frac{\Gamma (1-\tau)}{(-4q)^{1/2}} 
\]
\[
\times\left[
\Theta ( |x| - |y| ) {\cal W}_{\tau, 1/2}( (-4 q)^{1/2} |x|) 
{\cal M}_{\tau, 1/2}( (-4 q)^{1/2} |y|)  -
( x \leftrightarrow y) \right].
\]

The values $E$ for which ($H_D - E$)$^{-1}$  does not exist constitute the eigenvalues of
$H_D$, and they are obtained from the points for which the Gamma function $\Gamma(1 - \tau)$ is not defined, that is,
$1 -\tau$ is a negative integer number. By recalling the expressions of $p$, $q$ and
$\tau$, the condition
$1 - \tau = - n$, $n =  0,1, 2,3, \cdots$, gives
\[
E_n = - \frac{\kappa^2m}{2 \hbar^2}\, \frac{1}{n^2} \quad
n = 1, 2, 3,\cdots,
\]which coincide with the eigenvalues of the usual {\sc 3D} hydrogen atom model.

These eigenvalues are twofold degenerated and a basis $\{ \phi_{n,1}, \phi_{n,2} \}$ of the subsequent eigenspace
is 
\[
\phi_{n,k}(x) = \Theta((-1)^k x) \; {\cal W}_{\tau, 1/2} ((-4q)^{1/2}|x|),
\quad  k=1,2.
\]

The negative eigenvalues of other extensions $\dot H_{\hat U}$ are harder to get and numerical computation should be
employed. In the following particular cases of interest are selected.
Let $\Psi(x) := \frac{d}{dx}(\ln{\Gamma(x)})$ and define
\[
\omega(E) := \frac{2m\kappa}{\hbar^2}\left[\ln\left(\frac{\hbar^2}{2m}\tau\right) 
+ 2 \Psi(1) - \Psi(1-\tau)-1\right]-\frac{(-2Em)^{1/2}}{\hbar}
\] with  $[\Gamma(1-\tau)]^{-1}$ and $\pm \omega(E) [\Gamma(1-\tau)]^{-1}$ denoting, 
respectively, the values of the lateral limits
$ \lim_{x \to 0^\pm} {\cal W}_{\tau, 1/2} ((-4q)^{1/2}|x|)$ and
\[
\lim_{x \rightarrow 0^\pm} 
\left(\frac{d}{dx}{\cal W}_{\tau, 1/2}((-4q)^{1/2}|x|) \pm
\frac{2m\kappa}{\hbar^2} {\cal W}_{\tau, 1/2} ((-4q)^{1/2} |x|)
\ln( \pm \kappa x)\right).
\]
Given a unitary matrix $\hat U$, the candidates for eigenfunctions of $\dot H_{\hat U}$ must satisfy the corresponding
boundary conditions. 

\begin{exem2}\label{exemU1} Let's take  $\theta = \frac{\pi}{2}$, $a=1$ and $b=0$ so that
\[
\hat U = i
\left(\begin{array}{ll}
1  &  0 \\
0  &  1 
\end{array} \right),
\]
and the values of $E$ for which
$\omega (E) = -1$ are the eigenvalues of
$\dot H_{\hat U}$ and with multiplicity two; the corresponding eigenspace is spanned by
\[
\phi_k(x) = \Theta((-1)^k x) {\cal W}_{\tau, 1/2} ((-4q)^{1/2}|x|),
\quad k=1,2.
\] 
\end{exem2}

\begin{exem2}\label{exemU2}
Consider another case: $\theta = \frac{\pi}{2}$, $a=i$ and $b=0$ so that
\[
\hat U = i
\left(\begin{array}{ll}
i  &  0 \\
0  &  -i 
\end{array} \right)
\]
and the values of $E$ for which $\omega (E) = 0$ are found to be nondegenerate eigenvalues of
$\dot H_{\hat U}$, and for each eigenvalue the corresponding eigenspace is spanned by
\[
\phi(x) = \Theta(x) {\cal W}_{\tau, 1/2} ((-4q)^{1/2}|x|).
\]
\end{exem2}

These two cases illustrate that the behavior of eigenfunctions
are not related only to the parity of the potential $V_C(x)$, since  there are
cases for which the eigenfunctions do not have a definite parity and cases with eigenvalues simple as well as with
multiplicity two. These different possibilities are directly related to the singularity of the potential and depend
on the selected self-adjoint extension.

\subsection{Permeability of the origin}
Another question that has been discussed in the literature is about the permeability of the origin in the
unidimensional hydrogen atom; see, for instance \cite{And1,Lou,Kuras, Kuras2,Moshinsky,FLM2,RNewton}. Some authors
consider the origin an impermeable barrier, while others assume it is permeable. Again the answer strongly depends on
the self-adjoint extension considered (see also \cite{FLM2}), as illustrated ahead. Here the definition of permeability
is through the probability current density; for simplicity we assume $\hbar=1$ and $m=1$.

Recall that the probability current density $j(x)$ in {\sc 1D} is given by
\[
j(x) = \frac{i}{2} \left(\phi(x) \overline{\phi'(x)} - \phi'(x) \overline{\phi(x)} \right), 
\quad \phi \in \dom \dot H_{\hat U},
\]and it satisfies
the continuity equation
\[
 \frac{\partial}{\partial t} | \phi(t,x)|^2 + 
\frac{\partial}{\partial x} j(t,x) =0.
\] Our previous results in Subsection~\ref{subsectSae} show that $\lim_{x\to0^\pm}j(x)$ do exist (see ahead). 

For each  $\phi \in \dom \dot H_{\hat U}$, on integrating by parts we get
\[
0=\langle \dot H_{\hat U} \phi, \phi \rangle - 
\langle  \phi, \dot H_{\hat U} \phi \rangle =
i \lim_{\varepsilon \to 0}
[j(\varepsilon)- j(-\varepsilon)].
\]
Hence, the function $j(x)$ can be continuously defined at the origin $j(0)$ via lateral limits. 
Physically this relation means that the current density is isotropic at the origin, in spite of the strong
singularity there. A simple observation shows that
\[
j(x) = \im (\phi'(x) \overline{\phi(x)}),
\quad \phi \in \dom \dot H_{\hat U},
\] where $\im$ indicates imaginary part, so that
\[
j(0) =   \lim_{x \to 0^+} \im (\phi'(x) \overline{\phi(x)})
       =   \lim_{x \to 0^-} \im (\phi'(x)\overline{\phi(x)}).
\]

Since $\phi'(0^+)$ and $\phi'(0^-)$ can be divergent, we use using Lemma~\ref{lila} to obtain
\[
j(0)  =  \lim_{x \to 0^+} \im (\tilde{\phi}(x) \overline{\phi(x)} )
        =  \lim_{x \to 0^-} \im (\tilde{\phi}(x) \overline{\phi(x)}).
\]
Note that it is exactly this relation that guaranties that $j(0)$ is well defined and finite. 

We are now in position of giving a rigorous definition of permeability: If  $j(0)=0$, $\forall \phi \in \dom \dot
H_{\hat U}$, the electron is completely reflected when approaching the origin, and so we say the origin is not
permeable (or is impermeable), so that the regions $x<0$ and
$x>0$ are kept separated by the singularity. If
$j(0)\ne0$ we say the origin is permeable.

Next we study the current density related to $\dot H_{\hat U}$ in three cases.

\

\noindent
{\bf Case 1.}
$(I-\hat U)$ is invertible.  Since $A = -i(I-\hat U)^{-1}(I+\hat
U)$ is a self-adjoint matrix, the boundary conditions of $\dot H_{\hat U}$ become
\[
\left(\begin{array}{cc} 
\tilde{\phi}(0^+) \\
\tilde{\phi}(0^-)
          \end{array}\right)
=
\left(\begin{array}{cc}
u   &     z  \\
\overline{z}   &    v
\end{array}\right)
\left(\begin{array}{cc}
-\phi(0^+) \\
\phi(0^-)
          \end{array}\right),
\quad  u, v \in \R, z \in \C,
\]
and $u$, $z$ and $v$ are functions of the entries of  $\hat U$. 
The boundary conditions become
\begin{eqnarray}\label{denscorrecapar} 
\lim_{x \to 0^+} \tilde{\phi}(x) & = & - u \lim_{x \to 0^+} \phi(x) +
z \lim_{x \to 0^-} \phi(x) \\
\lim_{x \to 0^-} \tilde{\phi}(x) & = & - \overline{z} \lim_{x \to 0^+} \phi(x) +
v \lim_{x \to 0^-} \phi(x) 
\end{eqnarray}

Multiply the first equation (before taking limits) by 
$ \overline{\phi(x)}$ to get
\[
\lim_{x \to 0^+} \tilde{\phi}(x) \overline{\phi(x)}  =
-u \lim_{x \to 0^+} | \phi(x) |^2
+ z \lim_{x \to 0^+} \phi(-x) \overline{\phi(x)},
\]
and so
\[
j(0) = \lim_{x \to 0^+} {\rm Im}( \tilde{\phi}(x) \overline{\phi(x)}) =
\lim_{x \to 0^+} {\rm Im} (z  \phi(-x) \overline{\phi(x)}).
\]

Therefore, if $z = 0$ then $j(0)=0$,
$\forall \phi \in \dom \dot H_{\hat U}$, and we have found a family of self-adjoint extensions for which the origin is
not permeable. Example~\ref{exemU1} above corresponds to the self-adjoint matrix
\[
-i(I-\hat U)^{-1}(I+\hat U) =
\left(\begin{array}{cc}
1   &     0  \\
0   &     1
\end{array}\right),
\]
and the origin is not permeable in this case.
 
\

\noindent
{\bf Case 2.} $(I+\hat U)$ is invertible. The matrix
$A = i (I+\hat U)^{-1}(I-\hat U)$ is also self-adjoint and the boundary conditions of $\dot H_{\hat U}$ take the form
\[
\left(\begin{array}{cc}
u   &     z  \\
\overline{z}   &    v
\end{array}\right)
\left(\begin{array}{cc} 
\tilde{\phi}(0^+) \\
\tilde{\phi}(0^-)
          \end{array}\right)
=
\left(\begin{array}{cc}
-\phi(0^+) \\
\phi(0^-)
          \end{array}\right),
\quad  u,v \in \mathbb R, z \in \mathbb C,
\]
and $u$, $z$ and $v$ are functions of the entries of $\hat U$. Table~1 shows the current density at the origin for
various values of $u$, $z$ and $v$. 

\begin{center}
\def\tablename{Table}\label{tabelinha}
\begin{table}[htb!]
\caption{Current density for invertible $(I+\hat U)$.} \vspace*{0.1cm}
\centering
\begin{tabular}{|c|c|c|} \hline
$u$, $z$, $v$ $\neq 0$        &  
$j(0) =  {\rm Im} \left(- \frac{z}{u}  
\lim_{x \to 0^-}  \tilde{\phi}(x) \overline{\phi(-x)}  \right)$  
\\ \hline
$z = 0$               &
j(0) = 0
\\ \hline
$z \neq 0$ and $u = 0$               &
$j(0) =  {\rm Im} \left(- \frac{1}{z}  
\lim_{x \to 0^+}  \phi(x) \overline{\phi(-x)} \right)$
\\ \hline
$z \neq 0$ and $v = 0$               &
$j(0) = {\rm Im} \left( \frac{1}{\overline{z}}  
\lim_{x \to 0^-}  \phi(x) \overline{\phi(-x)} \right)$
\\ \hline \hline 
\end{tabular}
\end{table}
\end{center}

Note that Dirichlet boundary condition (so $\hat U=I$) is a particular case with $z=0$ (the matrix $A=0$), and since
this case can not be an extension of another self-adjoint extension of $\dot H$,  we conclude that the
current density vanishes for all
$\phi\in \dom \dot H_{\hat U}$ if, and only if, $z = 0$. In other words, if $(I+\hat U)$ is invertible, the origin is
impermeable precisely if $z=0$. As expected, Dirichlet boundary condition implies the origin is impermeable.

\

\noindent
{\bf Case 3.} Both $(I+\hat U)$ and $(I-\hat U)$ are not invertible. This case amounts to 
\[
\det(I+\hat U)=0=\det (I-\hat U),
\]
which turns out to be equivalent to the following matrix representation
\[
\hat U = \left(
\begin{array}{cc}
-u &  v \\
\overline{v} &  u
\end{array}\right), \quad 
u \in \mathbb R, v \in \mathbb C, \quad
|u|^2+|v|^2 = 1.
\]
The current density always vanishes at the  origin if, and only if, $v=0$, that is, the matrix $\hat U$ equals
\[
\left(
\begin{array}{cc}
-1 &  0 \\
0 &  1
\end{array}\right) \quad  \hbox{or}  \quad
\left(
\begin{array}{cc}
1 &  0 \\
0 &  -1
\end{array}\right).
\]

If $v \neq 0$, we have
\[
j(0) = {\rm Im}  \left(\frac{v}{1+u}\lim_{x \to 0^-}  \tilde{\phi}(x) \overline{\phi(-x)}  -
\frac{i v}{(1+u)} \lim_{x \to 0^-}  \phi(x) \overline{\phi(-x)} +
i \frac{1-u}{1+u} \lim_{x \to 0^+} | \phi(x)|^2 \right).
\]

\

\begin{exem2}\label{exemiiej}{\rm
Consider the self-adjoint extension $\dot H_{\hat U}$ which corresponds to the unitary matrix
\[
\hat U = i 
\left(
\begin{array}{ccc}
\sqrt2 /2  & -\sqrt2 /2  \\
\sqrt2 /2  & \sqrt2 /2 
\end{array}\right).
\]
The domain of $\dot H_{\hat U}$ constitutes of the  $\psi \in \dom \dot H^*$ so that
\[
\left(\begin{array}{cc} 
\tilde{\psi}(0^+) \\
\tilde{\psi}(0^-)
          \end{array}\right)
=
\left(\begin{array}{cc}
\sqrt 2  &     -i  \\
i        &   \sqrt 2
\end{array}\right)
\left(\begin{array}{cc}
-\psi(0^+) \\
\psi(0^-)
          \end{array}\right),
\]
and such conditions imply
\[
j(0) =   \lim_{x \to 0^+} \im (-i \psi(-x) \overline{\psi(x)}).
\]
The values of $E$ for which
$\omega(E)= - \sqrt 2$ are eigenvalues of $\dot H_{\hat U}$ of multiplicity two, and the corresponding eigenfunctions
are
\[
\phi_k(x) = \Theta((-1)^k x) {\cal W}_{\tau, 1/2} ((-4q)^{1/2}|x|),
\quad k=1,2.
\]
By taking the linear combination $\psi(x) = \phi_1(x) + \phi_2(x)$, and the asymptotic behavior of such
eigenfunctions near zero, discussed in Subsection~\ref{subsectNE1},
we obtain 
\[
j(0) =   -\Gamma(1 - \tau)^{-2} \neq 0,
\]
that is,  if the electron is in this eigenstate it is transmitted through the origin.
}
\end{exem2}

We conclude that there are extensions for which the origin is permeable and for others it is impermeable. Andrews 
\cite{And1} defines
$j(x) = i[\overline{F'(x)}F(x) - \overline{F(x)} F'(x)]$ but computes $j(0)$ only for eigenfunctions; it is clear that
$j$ vanishes if the eigenvalue is nondegenerate, since the corresponding eigenfunction can be taken real. Andrews
mentioned the possibility of zero current in case of degenerated eigenvalues. In our Example~\ref{exemU1} the
eigenvalues have multiplicity two and the current density is zero, whereas for the operator in Example~\ref{exemiiej}
the eigenvalues have multiplicity two and the origin is permeable; therefore both possibilities are allowed in case of
multiple eigenvalues.

We note that the  analysis of Moshinsky 
\cite{Moshinsky}, although interesting, considers the ``eigenfunctions''  ${\cal W}_{\lambda}(-z)$ and ${\cal
W}_{\lambda}(z)$  that do no belong to $L^2(\mathbb R \setminus\{0\})$.

\section{Potentials Via Laplace Equation}\label{PotLaplacesection}
This brief section deals with  Schr\"odinger operators with potentials $V$ along the fundamental solutions of Laplace
equation 
\[
\Delta V =0.
\]

As mentioned in the Introduction, in physics sometimes one assumes that in each dimension the potential describing the
Coulomb interaction is the fundamental solutions of Laplace equation \cite{Barton,Vladimirov};
recall that these solutions are (take $\kappa>0$)
\[
V_1(x) = \kappa |x|,\quad V_2(x)= \kappa \ln |x|,\quad V_3(x)=-\frac{\kappa}{|x|},
\]in {\sc 1D}, {\sc 2D}  and {\sc 3D}, respectively.

The case of $V_3$ in {\sc 3D} is standard and was recalled in Section~\ref{SAEsection3D}; we underline that the
operator
\[
{H}  =  - \frac{\hbar^2}{2m}  \Delta  + V_3(x), \quad \dom {H} = C_0^\infty(\mathbb R^3),
\]
is essentially self-adjoint and its unique self-adjoint extension has the same action but domain $\mathcal
H^2(\R^3)$; this extension has nonempty discrete and essential spectra.

The  case of potential $V_2$ was analyzed by Gesztesy and Pittner \cite{GesztesyPittner} and they state the following
result:

\begin{teor2}
The operator
\[
H = -\frac{\hbar^2}{2m} \Delta + \kappa \ln|x|, \quad 
\dom H = C_0^\infty(\mathbb R^2),
\]
is essentially self-adjoint, and its unique self-adjoint extension has empty essential spectrum.
\end{teor2}

Now we consider the unidimensional Schr\"odinger operator
\[
H  = - \frac{\hbar^2}{2m} \frac{d^2}{dx^2} + \kappa |x|, \quad 
\dom H = C_0^\infty (\mathbb R),
\]
whose adjoint $H^*$ has the same action as $H$ but domain
\[
\dom H^* = \left\{ \phi \in \LL^2(\R): \phi, \phi' \in \mathrm{AC}(\mathbb R), H^* \phi \in \LL^2(\mathbb R) \right\}.
\]Since $V_1$ is a bounded from below and continuous potential, with 
\[
\lim_{|x|\to\infty} V_1(x)=\infty,
\] the
following theorem follows from general results \cite{RS2,deOIST}.

\begin{teor2}\label{liinderedsi}
The above unidimensional operator $H$ is essentially self-adjoint, its unique self-adjoint extension $H^*$ is bounded
from below and has empty essential spectrum.
\end{teor2}

By solving the eigenvalue equation
\[
-\frac{\hbar^2}{2m} \phi'' + \left( \kappa |x| - E \right) \phi = 0
\]in terms of Airy functions we have found the eigenvalues $E_n$ are simple and with asymptotic behavior
\[
E_n\sim \frac{\hbar^2}{2m} 
\left[ \frac{m\kappa}{\hbar^2} \frac{3\pi}{4} (4n-3) \right]^{2/3},
\quad n \to \infty.
\]

Hence, for  Schr\"odinger operators $H$ with potentials along the fundamental solutions of the Laplace equation, we
have:

\begin{enumerate}
\item $H$ is essentially self-adjoint in  $C_0^\infty(\mathbb R^3)$ and its self-adjoint extension
has both nonempty discrete and essential spectra.

\item
 $H$ is essentially self-adjoint in $C_0^\infty(\mathbb R^n)$ and its self-adjoint extension has purely discrete
spectrum for $n=1,2$.
\end{enumerate}

However, for Schr\"odinger operators whose potential is the Coulomb one, i.e., $V_C(x)$,
we have:

\begin{enumerate}
\item
 The deficiency indices are equal to 0 in $C_ 0^\infty(\mathbb R^3)$.

\item
 The deficiency indices are equal to 1, 1 and 2 in $C_ 0^\infty(\mathbb R^3 \setminus \{0\})$, $C_ 0^\infty(\mathbb R^2
\setminus \{0\})$ and $C_ 0^\infty(\mathbb R \setminus \{0\})$, respectively.
\end{enumerate}

\section{Conclusions}\label{ConclusionsSection}

Although the {\sc3D} usual model hamiltonian $H$ with Coulomb potential $V_C$, $\dom H=C_0^\infty(\R^3)$, is essentially
self-adjoint, in $\R^n$ the $1/|x|$ singularity imposes the initial domain must be $C_0^\infty(\R^n\setminus\{0\})$,
$n=1,2$, and the corresponding operators $\dot H$ have deficiency indices equal to $2$ an $1$, respectively; hence with
infinitely many self-adjoint extensions. For the sake of comparison, we have also considered the origin removed in
$\R^3$, that is,
$\dot H$ with domain $C_0^\infty(\R^3\setminus\{0\})$: the deficiency indices are equal to 1 in this case. In each
case, all self-adjoint extensions have been found.

In {\sc1D} the question of permeability of the origin was analyzed and the answer depends strongly on the
self-adjoint extension considered. Due to particular examples discussed, we conclude that the multiplicity two of the
eigenvalues does not determine the permeability.

We have paid particular attention to the {\sc 1D} case, since there are many papers in the literature about this
model and occasionally with conflicting conclusions. We have found that these conflicting positions have been
originated from boundary conditions imposed mainly on ``physical basis'' that can fail for strong singularities, as is
the case of $V_C$ in one-dimension. We expect to have clarified the situation, and the next step could be
presenting  arguments to select the extension(s) to be considered natural, with the consequent implications as, for
instance, the  permeability of the origin.

Finally, we have found remarkable that, for potentials in $\R^n$, $n=1,2,3,$ given by
fundamental solutions of Laplace equation, the corresponding initial hermitian operators with domain
$C_0^\infty(\R^n)$ are always essentially self-adjoint.

 \subsubsection*{Acknowledgments} {\small AAV was  supported by CAPES (Brazil). CRdO acknowledges partial
support from CNPq (Brazil).}

\


\begin{thebibliography}{99}

\bibitem{JSS} R. V. Jensen, S. M. Susskind and M. M. Sanders, Phys. Rep. {\bf201}, 1 (1991). 

\bibitem{DKS} N. B. Delone, B. P. Krainov and D. L. Shepelyansky, Sov. Phys.-Usp. {\bf26}, 551 (1983). 

\bibitem{LCO3}A. L\'opez-Castillo  and C. R. de Oliveira,  Chaos Sol. Fract. {\bf 15}, 859 (2003).

\bibitem{CC} M. W. Cole  and M. H. Cohen, Phys. Rev. Lett. {\bf 23}, 1238 (1969).

\bibitem{Wong} C. M. Wong, J. D.  McNeill, K. J. Gaffney, N.-H. Ge, A. D. Miller, S. H.  Liu and
C. B. Harris, J. Phys. Chem. B {\bf 103}, 282  (1999).

\bibitem{VRs} V. S. Vrkljan,  Zeits.  f. Physik
{\bf 52}, 735 (1928).   Zeits. f. Physik {\bf 54}, 133 (1929).

\bibitem{Lou} R. Loudon, Amer. J. Phys. {\bf 27}, 649 (1959).

\bibitem{And1}M. Andrews,  Amer. J. Phys. {\bf 34}, 1194 (1966).

\bibitem{HR} L. K. Haines and D. H. Roberts,  Amer. J. Phys. {\bf 37}, 1145 (1969).


\bibitem{And2} M. Andrews,  Amer. J. Phys. {\bf 44}, 1064 (1976).

\bibitem{GZ}  J. F. Gomes  and  A. H. Zimerman, Amer. J. Phys. {\bf 48}, 579 (1980).

\bibitem{SL} H. N. Spector  and J. Lee, Amer. J. Phys. {\bf 53}, 248 (1985).

\bibitem{DPST}  L. S. Davtyan, G. S.  Pogosyan,  A. N. Sissakian  and V. M. Ter-Antonyan, J.
Phys. A: Math. Gen. {\bf 20}, 2765 (1987).


\bibitem{BKB} L. J. Boya, M.  Kmiecik  and A.  Bohm,  Phys. Rev. A {\bf 37}, 3567 (1988).


\bibitem{NVS} H. N. N\'u\~nes-Y\'epez, C. A.  Vargas  and A. L. Salas-Brito, Phys. Rev. A {\bf 39}, 4306  (1989).

\bibitem{LC} W.-C. Liu  and C. W. Clark, J. Phys. B: At. Mol. Opt. Phys. {\bf 25}, L517 (1992).

\bibitem{OL} U. Oseguera  and M. de Llano, J. Math. Phys. {\bf 34}, 4575 (1993).

\bibitem{FLM} W. Fischer, H. Leschke and P. M\"uller,  J. Math.  Phys. {\bf 36}, 2313 (1995).

\bibitem{XDD}D. Xianxi,  J. Dai  and J. Dai,  Phys. Rev. A {\bf 55}, 2617 (2001).

\bibitem{LL} Q.-S. Li  and J. Lu, Chem. Phys.  Lett., {\bf 336}, 118 (2001).

\bibitem{LCO6} A. L\'opez-Castillo  and C. R. de Oliveira,  J. Phys. A: Math. Gen. {\bf39}, 3447 (2006). 

\bibitem{CRdO} C. R. de Oliveira, Boundary triples, Sobolev traces and self-adjoint extensions in multiply
connected domains. Submitted for publication.

\bibitem{RS2}  M. Reed and B. Simon, {\it Fourier analysis, self-adjointness} (Academic Press, New York, 1975).

\bibitem{deOIST} C. R. de Oliveira, {\it Intermediate spectral theory and quantum dynamics} (Birkh\"auser, Basel), to
appear.

\bibitem{Msh} C. M\"uller, {\it Spherical Harmonics}, LNM {\bf17} (Springer-Verlag, Berlin, 1966).

\bibitem{AJS} W. O. Amrein, J.  M. Jauch and K. B. Sinha,  {\it Scattering theory in quantum mechanics. Physical
principles and mathematical methods} (Benjamim, London, 1977).

\bibitem{Kuras}P. Kurasov, J. Phys. A: Math. Gen. {\bf 29}, 1767 (1996).

\bibitem{Kuras2} P. Kurasov,  J. Phys. A: Math. Gen. {\bf30}, 5583 (1997).

\bibitem{Moshinsky} M. Moshinsky, J. Phys. A: Math. Gen. {\bf 26}, 2445 (1993).

\bibitem{Abro} M. Abramowitz and I. A. Stegun,  
{\it Handbook of  Mathematical Functions} (Dover, New York, 1972).

\bibitem{Grad}L. S. Gradshteyn and I. M. Ryzhik, {\it Table of Integrals, Series, and Products}
(Academic Press, San Diego, 1994).

\bibitem{RS1} M. Reed and B. Simon, {\it Functional analysis.} Second edition (Academic Press, New York, 1980).

\bibitem{FLM2}  W. Fischer, H. Leschke and P. M\"uller, J.  Phys. A {\bf 30}, 5579 (1997).

\bibitem{RNewton} R. G. Newton,  J. Phys. A: Math. Gen. {\bf27}, 4717 (1984).

\bibitem{Barton}G. Barton, {\it Elements of Green's Functions and Propagation: Potentials, Diffusion and Waves} (Oxford
Science Publications, Oxford, 1989).

\bibitem{Vladimirov}V. S. Vladimirov, {\it Equations of Mathematical Physics}  (Mir, Moscow, 1984).

\bibitem{GesztesyPittner} F. Gesztesy and L. Pittner,  J. Phys A: Math. Gen. {\bf 11}, 679 (1978).




\end{thebibliography}
\end{document}